\documentclass[sigconf]{acmart}

\AtBeginDocument{%
  \providecommand\BibTeX{{%
    \normalfont B\kern-0.5em{\scshape i\kern-0.25em b}\kern-0.8em\TeX}}}

\copyrightyear{2022} 
\acmYear{2022} 
\setcopyright{acmcopyright}\acmConference[FDG '22]{FDG '22: Proceedings of the
17th International Conference on the Foundations of Digital Games}{September 5--8,
2022}{Athens, Greece}
\acmBooktitle{FDG '22: Proceedings of the 17th International Conference on the
Foundations of Digital Games (FDG '22), September 5--8, 2022, Athens, Greece}
\acmPrice{15.00}
\acmDOI{10.1145/3555858.3555883}
\acmISBN{978-1-4503-9795-7/22/09}

\usepackage{multirow}

\usepackage{graphicx}  
\usepackage{subfigure}    
\usepackage{float}     
\usepackage{booktabs} 
\usepackage{tabularx}
\usepackage{engord}
\usepackage{makecell}

\begin{document}

\title{Towards Understanding Player Behavior in Blockchain Games: \\A Case Study of Aavegotchi}

\author{Yu Jiang}
\affiliation{%
  \institution{The Chinese University of Hong Kong, Shenzhen}
  \city{Shenzhen}
  \country{China}}
\email{yujiang1@link.cuhk.edu.cn}

\author{Tian Min}
\affiliation{%
  \institution{The Chinese University of Hong Kong, Shenzhen}
  \city{Shenzhen}
  \country{China}}
\email{tianmin@link.cuhk.edu.cn}

\author{Sizheng Fan}
\affiliation{%
  \institution{The Chinese University of Hong Kong, Shenzhen}
  \city{Shenzhen}
  \country{China}}
\email{sizhengfan@link.cuhk.edu.cn}

\author{Rongqi Tao}
\affiliation{%
  \institution{y2z Ventures}
  \city{Shanghai}
  \country{China}}
\email{taorongqi@gmail.com}

\author{Wei Cai}
\authornote{Wei Cai is the corresponding author (caiwei@cuhk.edu.cn).}
\affiliation{%
  \institution{The Chinese University of Hong Kong, Shenzhen}
  \city{Shenzhen}
  \country{China}}
\email{caiwei@cuhk.edu.cn}


\begin{abstract}
Blockchain games introduce unique gameplay and incentive mechanisms by allowing players to be rewarded with in-game assets or tokens through financial activities. However, most blockchain games are not comparable to traditional games in terms of lifespan and player engagement. In this paper, we try to see the big picture in a small way to explore and determine the impact of gameplay and financial factors on player behavior in blockchain games. Taking Aavegotchi as an example, we collect one year of operation data to build player profiles. We perform an in-depth analysis of player behavior from the macroscopic data and apply an unsupervised clustering method to distinguish the attraction of the gameplay and incentives. 
Our results reveal that the whole game is held up by a small number of players with high-frequent interaction or vast amounts of funds invested. Financial incentives are indispensable for blockchain games for they provide attraction and optional ways for players to engage with the game. However, financial services are tightly linked to the free market. The game will face an irreversible loss of players when the market experiences depression. For blockchain games, well-designed gameplay should be the fundamental basis for the long-lasting retention of players.


\end{abstract}


\maketitle

\section{Introduction} \label{sec:intro}


Blockchain game is not a new term. As early as 2017, Cryptokitties\footnote{https://www.cryptokitties.co}, one of the first blockchain games, caused a sensation upon its release, accounting for more than 10\% of the total traffic on Ethereum. However, a large number of players quickly forwent Cryptokitties within one month. The possible reasons behind the rapid decline in the game's popularity include the oversupply of kitties, the 
reduction of player income, a widening gap among players, and the limitations of the current blockchain systems \cite{jiang2021cryptokitties}.

Fortunately, the progress of blockchain technology and the diversity of its ecosystem bring the possibility of further development of blockchain games. The first promotion is \textit{High-performance Infrastructure}: Sidechains and layer 2 enable credible, fast, low-cost, and high-frequency transactions for players. For example, the transactions per second (TPS) of Polygon\footnote{Polygon is a decentralized Ethereum scaling platform that enables developers to build scalable user-friendly DApps with low transaction fees without ever sacrificing on security.} is 65,000, which is 4000 times faster than Ethereum \cite{christodorescu2021universal}.
Another promotion is \textit{A Thriving Ecosystem}. The boom of specific blockchain projects has dramatically contributed to the growth of blockchain games. For example, decentralized exchanges (DEXs) \cite{zhang2022economic} can meet players' demand to exchange tokens; decentralized autonomous organizations (DAOs) can lower the barrier to participating in blockchain games by providing players with training sessions and lending services of in-game non-fungible tokens (NFTs) \cite{gudgeon2020defi}. These factors have led to the rise of blockchain games in 2020. According to DappRadar\footnote{https://dappradar.com/}, the number of unique active wallets (UAW) connected to decentralized applications (DApps) of games has reached 754,000, accounting for approximately 50\% of blockchain industry interactions \cite{Herrera2021bga}. The proliferation of players has also triggered a creative impetus among developers, with the number of games on the blockchain exceeding 1,100 by the end of 2021.

Why have blockchain games achieved such massive success in a short time? The main reason behind this is the blockchain, as an infrastructure, provides different gameplay and ecosystem from traditional games: 1) \textit{Diverse Ways of Participation}. With the ability for players to truly own their assets without relying on a centralized service provider, the reusability of digital assets across games and user-generated content offers a plethora of innovative gameplay options. 2) \textit{Open Economy System}. Blockchain games can issue their own fungible tokens to build in-game economies, giving developers a richer and more creative space and the possibility of connecting their games with the whole ecosystem. Players can exchange in-game tokens without hindrance for USDT, USDC, or ETH.
3) \textit{New Incentive Mechanism}. In addition to earning tokens by playing the game, the player can also gain in-game rewards, including assets and tokens by staking, which is a typical financial activity. Specifically, this can be interpreted as financing from developers to players, where developers use the financed cryptocurrencies to develop better game content and give back to players with assets or tokens as interest, which could create a better community.

However, the current blockchain games are still in a preliminary stage. Many games have a lifespan of just a few months or even weeks. We summarize the following open research questions to explore the problems existing in the development of blockchain games: 

\begin{itemize}
    \item[\textbf{Q1.}] What’s the trend of player activity in blockchain games?
    \item[\textbf{Q2.}] What attracts players to blockchain games, and why do they leave?
    \item[\textbf{Q3.}] What should the design of blockchain games focus on?
\end{itemize}

A straightforward way to answer these questions is to survey how blockchain games appeal to players. However, conducting user surveys will incur higher costs, especially on the blockchain. For the survey results to be representative, a relatively large number of player profiles would need to be collected. It can be challenging to send questionnaires to eligible players by their wallet addresses or select players who participated in certain blockchain games from the blockchain community. At the same time, the cryptocurrency used to pay as the rewards will entail a substantial monetary cost. Based on these considerations, it's wiser to extract the information we need from open-source data.

Hence, in this paper, we select a particular blockchain game, Aavegotchi, as the case and collect on-chain open-source data. The rationale support behind this choice can be summarized as follows: 1) Aavegotchi is the most popular fostering game on Polygon. 2) Aavegotchi is the first blockchain game with NFT generated by financial asset collateral. 3) Aavegotchi has experienced a relatively complete game life cycle from prosperity to recession. 
In the following content, we collect a year of player operation data from Aavegotchi. Then, we perform and visualize an in-depth analysis from the aspect of gaming and financial activities. We analyze the number of daily active addresses, functions called by users, and density distribution to reflect the player behaviors. 
Besides, we apply an unsupervised Self-Organizing Map (SOM) algorithm to divide user groups and discuss the player behavior in different clusters.
The main contributions of this work can be summarized as follows:
\begin{itemize}
    \item To the best of our knowledge, we are the first to quantitatively analyze player behavior in blockchain games from the perspective of finance and gameplay. Through the cluster results of players, we conclude that in-game staking can, to a certain extent, quickly attract players in the short term but cannot retain them in a long time.
    
    \item We point out the future direction of blockchain games.
    Financial incentives can sometimes create the illusion of game prosperity. As interest declines, players can withdraw from the game at any time.
    Blockchain games should be more cautious in utilizing the economic attributes and free market of blockchain. Instead, they should develop innovative gameplay taking advantage of the data sharing and transparency of the blockchain.
\end{itemize}

\section{Related Work}
\subsection{Blockchain Technology}
The blockchain has become the most disruptive technology since it was first mentioned in the Bitcoin whitepaper \cite{nakamoto2008bitcoin} released by Satoshi Nakamoto in 2008. The primitive blockchain is a distributed ledger system of all transactions across a peer-to-peer network that provides tamper-proof and traceable functions. With the further development of blockchain technology, Ethereum, known as the blockchain 2.0 platform, emerged \cite{wood2014ethereum}. In addition to the classic application of distributed ledger, the smart contracts on Ethereum are open-source programs that can be automatically executed without any centralized control \cite{9223754, fan2022mobile}, which allows the developers to create decentralized applications, covering the areas such as finance, and gaming \cite{a16z2021web3,8466786}.

Hence, many researchers have begun to focus on the study of DApps. Cai \emph{et al.} \cite{WeiCaiWEHFL2018} surveyed the state-of-the-art DApps to reveal the direction of blockchain development. 
A study in \cite{duan2021metaverse} illustrated the blockchain-driven metaverse and highlighted the representative applications for social good.
The work in \cite{min2022portrait} attempted to profile DApp users through publicly available data and applied an unsupervised clustering method to distinguish investors and players.
Moreover, some studies focus on niche areas in the gaming industry. For example, Min \emph{et al.} \cite{8848111} delved into blockchain games and analyzed the trends of blockchain games from a statistical approach. 
The work in \cite{jiang2022economic} analyzes the loot box trading market in blockchain games from the perspective of game theory.
However, blockchain games still face many problems, such as security and economic issues. The work in \cite{8811555} analyzed the blockchain game architecture and summarized a security overview from the perspective of the web server and smart contract, respectively.

\subsection{Behavioral Profiling}
The game industry has widely adopted the evaluation and visualization of player behavior.
Behavioral profiling highlights behavior patterns by condensing high-dimensional and high-volume data into descriptions \cite{wauck2020data, ahmad2019modeling, kleinman2020and}.
To explore what is behind a behavioral pattern, quantitative approaches, such as deep learning, data mining, and statistical analysis, effectively identify and analyze patterns from players' operation data.
The paper \cite{aung2019trails} applied cluster analysis to profile players' behavior in Just Cause 2, an open-world game.  
A study \cite{eaton2018attack} researched the impact of role familiarity on team performance in the Multi-player Online Battle Arena gaming environment using Analysis of Variance (ANOVA) and visualization techniques.
The work in \cite{ong2015player} found the optimal team composition for multiplayer online games by clustering player behavior.
Most researchers study players' behavior in traditional games, while only few works focuses on behavioral profiling in blockchain games.



\section{Introduction to Aavegotchi}
\subsection{Game Rules}
Aavegotchi is a simulation game (SLG) of crypto assets.
Like Tamagotchi\footnote{Tamagotchi is the world's first virtual pet, first released by Bandai in 1996.} introduced the world to digital pets, Aavegotchi introduces the world to playable NFTs, backed with a value of digitalization and medialization. 
Aavegotchi (can be called Gotchi\footnote{The game's name is the same as the ghost's name. For better presentation, Aavegotchi stands for the name of the game, and Gotchi stands for the name of the ghost below.}) is a kind of ghost in the form of a pixel living on the Polygon, whose nature is an NFT created by ERC-721 standard\footnote{ERC-721 is a free, open standard that describes how to build non-fungible or unique tokens on the Ethereum blockchain. Besides, ERC-721 tokens are all distinct.}.
Before raising the pixelated ghost baby, the player must summon it via a portal. There are three ways to acquire portals: in an auction, in a Drop Ticket activity, and at the Baazaar (Aavegotchi's secondary marketplace), among which auctions and Drop Ticket activity are time-limited events. If players missed these events, the only way left to acquire portals is through Baazaar. Once opening a portal, players can choose 1 of 10 Gotchis to summon. This choice often depends on the value of the Gotchi, which can be decided from intrinsic and rarity aspects.

\textbf{Intrinsic Value}: To claim a Gotchi from the portal, the players need to stake the required amount of aTokens\footnote{Aave is a decentralized lending protocol that allows players to earn interest on deposits or borrow assets. When depositing money in the Aave, the player will receive a corresponding amount of aToken, and aTokens accrues interest directly to the wallet.} interest-generating tokens from the Aave, as collaterals called Spirit Force. ATokens generate yield via Aave's LendingPool, which increases the number of aTokens held in the wallet. Therefore, the number of aTokens held in Gotchi's escrow address will increase over time, improving the intrinsic value of Gotchi.

\textbf{Rarity Value}: Upon summoning, the Gotchi will be given a rarity score based on the initial trait, which is randomly generated, but the scope of randomness depends on the Spirit Force. Then, the players can improve Gotchi's rarity value through various activities, which can be understood as the main gameplay.
\begin{figure}[htbp]
	\centering
    \includegraphics[width=1\columnwidth]{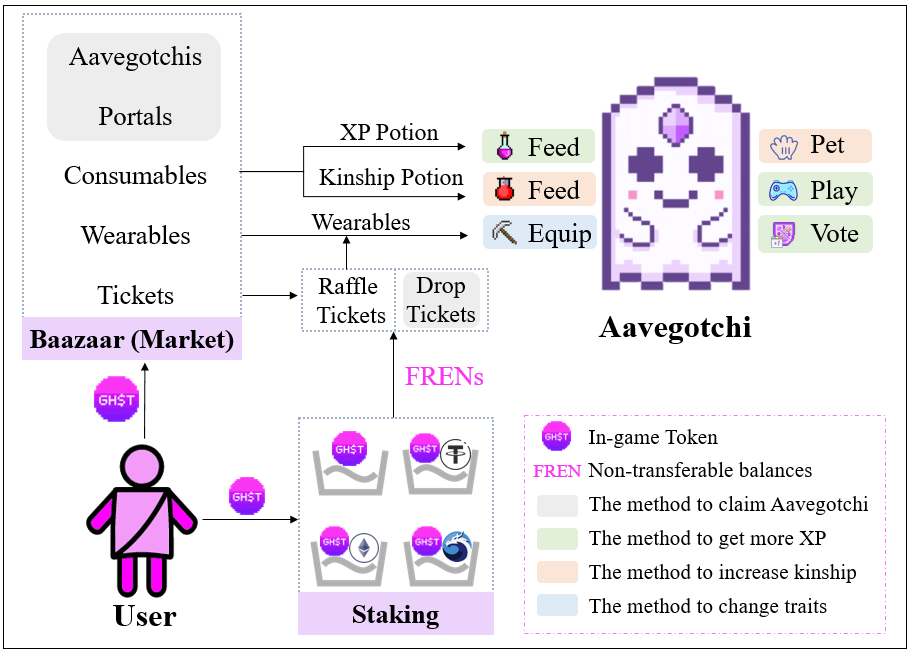}
    \caption{Game Framework}
    \label{fig:framework}
\end{figure} 

As shown in Figure \ref{fig:framework}, the game framework has pointed out several ways to cultivate the Gotchi by improving the rarity value. Specifically, there are three main methods to improve the rarity score: increasing kinship, getting more experience (XP), and changing traits. Firstly, to increase the kinship, players can pet the Gotchis every twelve hours and feed the Gotchis with kinship potions bought from Baazaar using GHST, the native token launched by Aavegotchi. Then, players can play mini-games with Gotchis, participate in the voting activities held by AavegotchiDAO, and feed XP potion to Gotchis to get more XP. Besides, the traits of the Gotchi can be changed by equipping the wearables, which can be acquired from the Baazaar or using Raffle Tickets.

In the process of cultivation, players can earn money. When players participate in voting activities held by AavegotchiDAO and mini-games, they have the chance to be rewarded with GHST tokens. Besides, when players raise the rarity of their Gotchis to a higher level, they can win GHST rewards in rarity farming, a competition organized by the project. Moreover, the Gotchi with high rarity can also be sold at a high price in the secondary market.

\subsection{In-game Token Staking}
How to get drop tickets and raffle tickets? In this part, we will describe the most innovative and exciting aspect of the game. Staking is the process of locking up cryptocurrency in exchange for rewards. GHST can be staked to earn FRENs, the non-transferable balances in Aavegotchi. Currently, Aavegotchi offers four types of staking: staking GHST, staking GHST-QUICK, staking GHST-USDC, and staking GHST-WETH. 
FRENs earned by players through staking any single or a pair of the above tokens will enable them to redeem Drop and Raffle Tickets.

\subsection{The Timeline of Aavegotchi Development}
\begin{figure}[htbp]
	\centering
    \includegraphics[width=1\columnwidth]{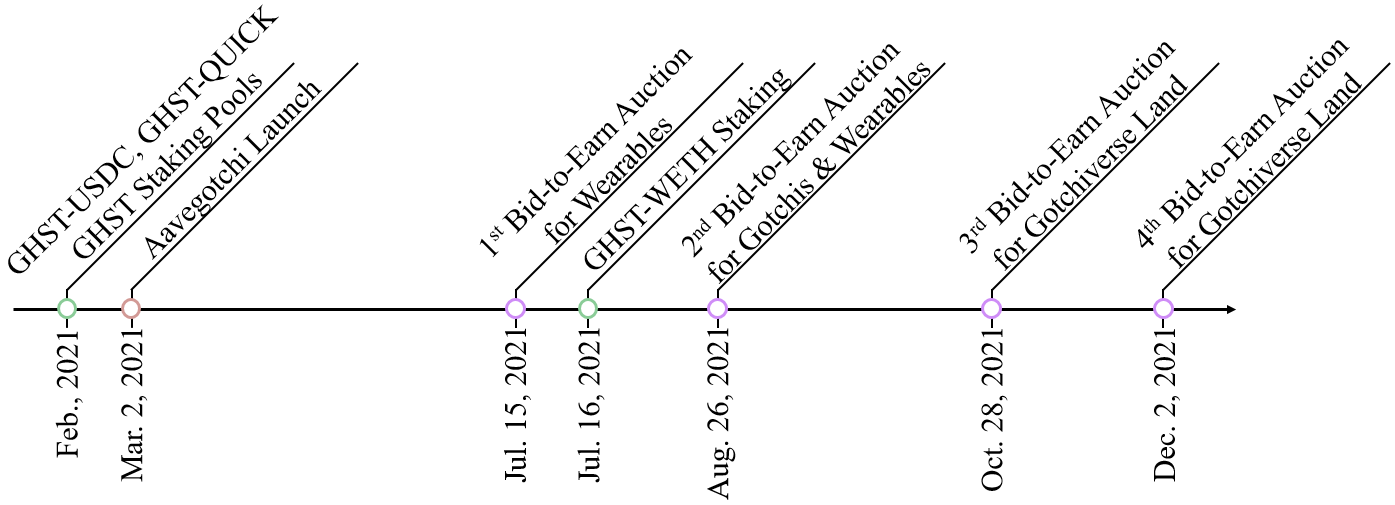}
    \caption{The Timeline of Aavegotchi Development}
    \label{fig:big_event}
\end{figure} 
From Figure \ref{fig:big_event}, we can observe that the three staking pools, staking GHST, staking GHST-QUICK, and staking GHST-USDC, were launched in February 2021, which were nearly one month before the official release of gameplay.
Since its launch, Aavegotchi has posted many extraordinary events. In this subsection, we will present the influential events in chronological order, shown in Figure \ref{fig:big_event}.
On March 2, 2021, Aavegotchi was released and followed by a so-called bid-to-earn auction, the most distinctive event ever held by Aavegotchi. A bid-to-earn auction is a typical \emph{English Auction} in which the opening bid starts low and increases as buyers bid for the item until one buyer is willing to pay a certain amount and a higher bid isn't received during the given time. Every time a participant is outbid, that person earns a payout up to a percentage of his original bid. The bid-to-earn auctions motivate the players to keep searching for earnings.

The bid-to-earn auction was held four times in 2021 on July 15 (1\textsuperscript{st} bid-to-earn auction), August 26 (2\textsuperscript{nd} bid-to-earn auction), October 28 (3\textsuperscript{rd} bid-to-earn auction) and December 2 (4\textsuperscript{th} bid-to-earn auction), each lasts for three days. The 1\textsuperscript{st} bid-to-earn auction is on a smaller scale for wearables. The 2\textsuperscript{nd} bid-to-earn auction was expanded to allow players to bid for portals as well as wearables.
When it comes to the release of Gotchiverse Realm\footnote{Gotchiverse Realm mixes elements of real-time strategy and sandbox farming with playable NFTs to deliver. Because Gotchiverse Realm was not fully developed at the time of writing, we will not cover it here.} project's release, the subsequent 3\textsuperscript{rd} and 4\textsuperscript{th} bid-to-earn auction were held for land in Gotchiverse.
Finally, it's worth noting that GHST-ETH liquidity rewards were launched on July 16, 2021.
\begin{figure}[htbp]
	\centering
    \includegraphics[width=0.8\columnwidth]{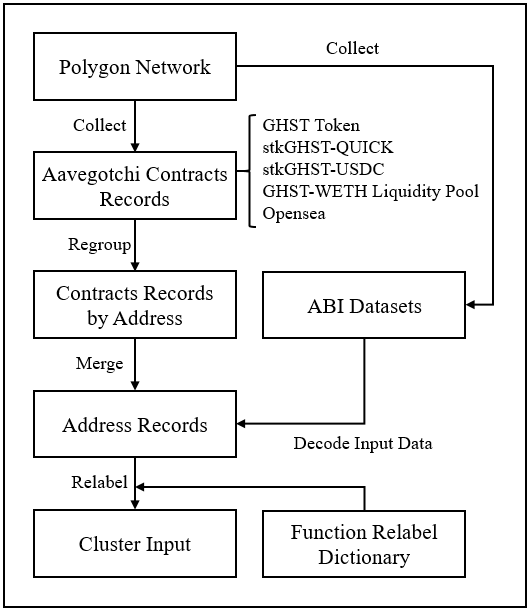}
    \caption{Data Collection Process}
    \label{fig:data_collection}
\end{figure} 
\section{Data Collection}
\subsection{Dataset}
As shown in Figure \ref{fig:data_collection}, we obtain the open records of addresses that have interacted with Aavegotchi's contracts from February 2, 2021, to February 25, 2022, with Polygonscan\footnote{https://polygonscan.com/}. We collect over 3,241,236 transaction records for five contracts through Aavegotchi's contract address. By reorganizing and consolidating the records of these contracts, we collect a record in terms of the individual addresses with which it had interacted. At the same time, we obtain the application binary interface (ABI) of the contracts from Polygon network and use it to decode the input parameters in the records to generate function names and parameters. After that, we merged the records with the same address to get a complete record of user transactions, including 31,464 independent addresses. Finally, we map a \emph{Function Relabel Dictionary} to the user records to transform the wide range of function names into functionally categorized labels. After normalizing and filtering out invalid transactions, we have a time series of all function calls by the user and use it as a preliminary material for the cluster input.



\subsection{Smart Contracts} \label{sec:smart contract}
In this paper, we collect five smart contracts related to Aavegotchi's major services from its official Wiki\footnote{https://wiki.aavegotchi.com/en/contracts}. The main functions of these five smart contracts can be introduced as:
\textbf{Opensea (for Aavegotchis and Wearables) Address} is responsible for most gameplay-related functions, including interacting with Gotchis, summoning new Gotchis, using consumables, managing equipment, and trading with other players. In addition, it enables players to handle miscellaneous tasks such as downloading or updating SVG images, generating URLs, etc.
\textbf{GHST Token Address} is able to record all token transaction information.
\textbf{stkGHST-QUICK Address, stkGHST-USDC Address, and GHST-WETH Liquidity Pool Address} enable players to stake single token of GHST, as well as token pairs, including GHST-QUICK, GHST-QUICK and GHST-WETH, to obtain FRENs.

\begin{table}[h]
  \caption{Basic Transactions and Address Statistics for Aavegotchi's Smart Contracts}
  \label{tab:smart_contracts_basic}
  \begin{tabular}{ccc}
    \toprule
    Smart Contract & Transaction & Unique Address\\
    \midrule
    Opensea & 2,720,415 & 12,135\\
    GHST Token & 117,375 & 29,937\\
    stkGHST-QUICK & 396,477 & 15,622\\
    stkGHST-USDC & 3,920 & 1,419\\
    GHST-WETH Liquidity Pool & 3,049 & 2,311\\
  \bottomrule
\end{tabular}
\end{table}
Table \ref{tab:smart_contracts_basic} shows the basic transactions and address statistics for five major smart contracts mentioned above. The data contained 31,464 individual addresses and generated a total of 3,241,236 transaction records. We can see that the Opensea contract, which carries the most significant functionality, has the highest number of transactions, but the number of independent addresses is less than the GHST Token contract. This can be explained by the fact that in the pre-publication phase of the game, most pre-purchasers involved in GHST Token were not actually active in the real game content. 

Including stkGHST-QUICK Address, stkGHST-USDC Address, GHST-WETH Liquidity Pool Address, the contracts used to provide staking services account for a substantial 12.45\% of the total trading volume, which shows that even as a DApp with a gaming-based business, its financial services can still attract a significant number of users targeting in-game rewards.

\subsection{Functions}
\begin{figure}[htbp]
	\centering
    \includegraphics[width=1\columnwidth]{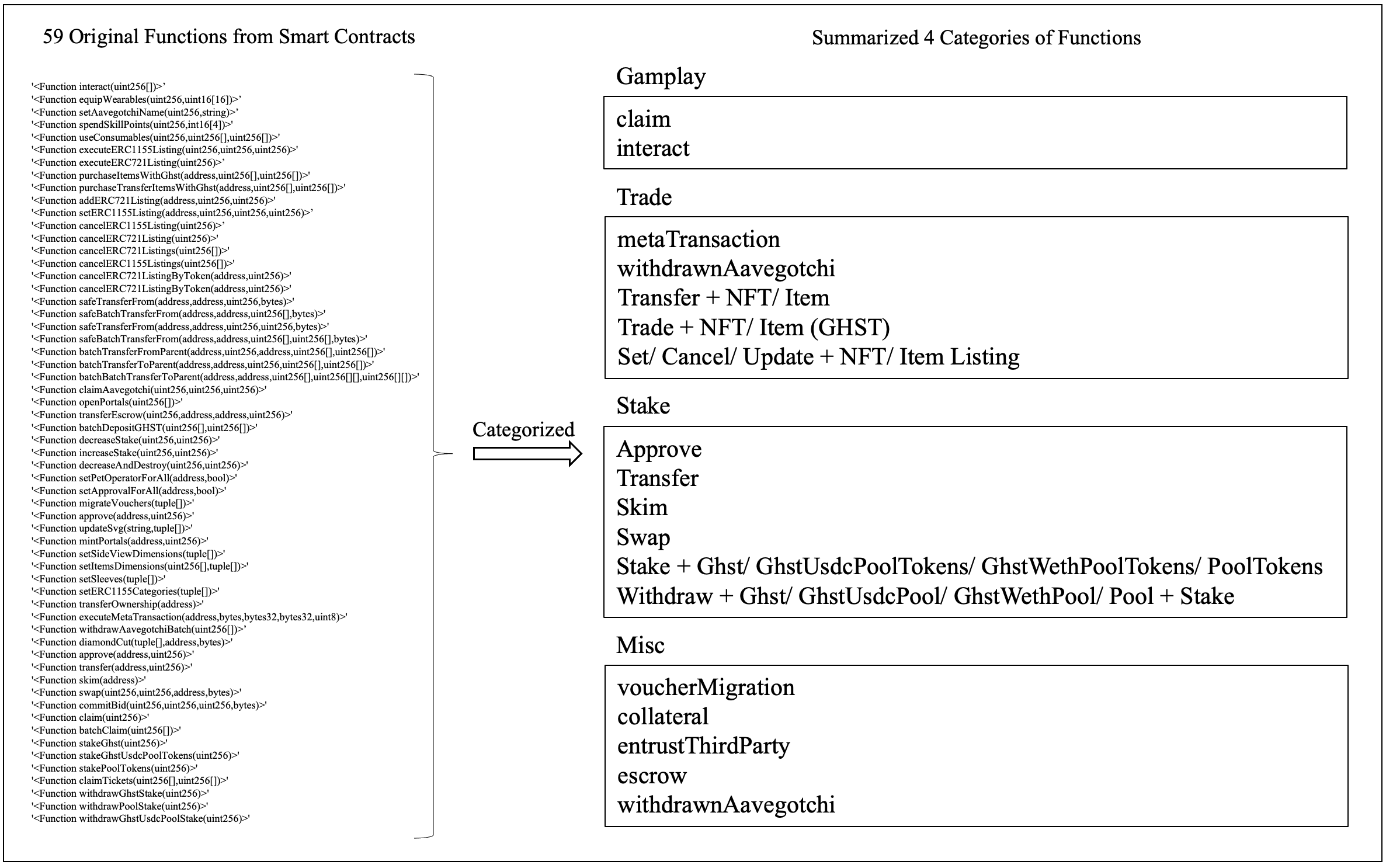}
    \caption{Function Relabel \& Categories}
    \label{fig:function_names}
\end{figure}

We found a total of 59 functions from the five main contracts that had been invoked, in which the Opensea contract contains the majority of functions, while most of the staking contracts include only a few functions that allow users to manipulate their funds. As depicted in Figure \ref{fig:function_names}, we regrouped 59 functions into 4 categories and created a relabel dictionary for translating function names, which is on the principle of classifying functions by the services they provide. The categories can be described as follows.

\textbf{Gameplay} contains the functions needed for the user to participate in the game itself, including summoning Gotchis and interacting with it.
\textbf{Trade} allows players to trade assets, including Gotchis, wearables and other items. They also include support for pending order trade and batch trade systems.
\textbf{Stake} provides an interface for staking and withdrawal of single coins as well as token pairs.
\textbf{Misc} includes functions for players to handle miscellaneous needs such as redeeming vouchers and transferring digital assets to other blockchain networks such as Ethereum. In addition, a number of administrative functions used by game developers are included.

\section{Methodology}
In this section, we present how we carry out feature extraction. Then, we describe how to apply SOM, an unsupervised clustering algorithm, to capture user groups with similar behavioral patterns.

\subsection{Feature Extraction}
From the description of smart contracts in Section \ref{sec:smart contract}, we can see that the operations related to gameplay are included in the Opensea contract, while the operations related to staking are included in the other four contracts. Hence, we divide user operations into interaction and staking, then work with the corresponding contracts to extract features.
\begin{figure}[htbp]
	\centering
    \includegraphics[width=1\linewidth]{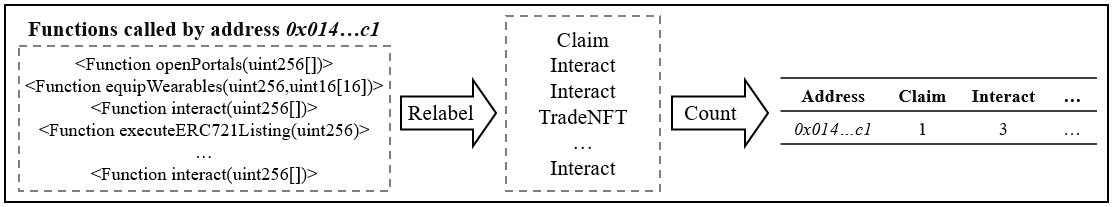}
    \caption{The Process of Interaction Operation Extraction}
    \label{fig:Interaction_Extraction}
\end{figure} 

For interaction data construction, we need to construct the operations performed in the Opensea contract for each address and count the times the functions were invoked. Figure \ref{fig:Interaction_Extraction} illustrates the process of interaction operation data construction for address \emph{0x014…c1}.
Firstly, we extract the function calls of each address from the Opensea contract and rearrange them in the form of time series to construct operation flow. For example, the address \emph{0x014…c1} firstly called '<Function openPortals(uint256[])>' to open the portals to acquire a Gotchi, then called '<Function equipWearables(uint256, uint16[16])>' to equip a wearable with his Gotchi, and called '<Function interact(uint256[])>' to pet his Gotchi, etc.
According the function labels mentioned in Figure \ref{fig:function_names}, we rename each function and abandon the labels in Misc category. Hence, the address \emph{0x014…c1} has his reconstructed flow of operations: \emph{'Claim', 'Interact', 'Interact', ...} Finally, we count the number of each label invoked by each user. After these processes, the transaction history of address \emph{0x014…c1} has been recorded as 1 'Claim', 3 'Interaction' and so on, as demonstrated in Figure \ref{fig:Interaction_Extraction}.

\begin{figure}[htbp]
	\centering
    \includegraphics[width=0.5\columnwidth]{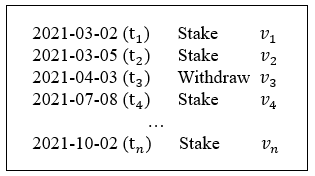}
    \caption{The Staking Operation of Address \emph{0x014…c1}}
    \label{fig:staking_operation}
\end{figure}
To format the staking operation, we need to calculate the weighted total value locked (TVL) for each address in the four staking contracts. From the staking transaction log, we can locate the number of token pairs acquired or given away by players when adjusting liquidity based on the hash value of transactions, which was then converted to USD at a price on the day\footnote{We have taken the price at the time of transactions made. We believe this value can better reflect players' decisions and behaviors and is the most feasible and persuasive solution to make the data discrete.} the transaction was made based on GHST price lists obtained from CoinGecko\footnote{https://www.coingecko.com/}. 
As shown in Figure \ref{fig:staking_operation}, we take the address \emph{0x014…c1} as an example, using $v_i$ and $t_i$ to denote the value and time of the $i$th staking operation of an address. This address has staked $v_1$ at $t_1$, staked $v_2$ at $t_2$, withdrawn $v_3$ at $t_3$, ..., and staked $v_n$ at $t_n$.
We can format the value and time sequences as $[v_1, v_2, v_3, \cdots,v_n]$ and $[t_1, t_2, t_3, \cdots, t_n]$. When the $i$th of operation of the address is to withdraw the fund, $v_i < 0$.
To further format a valid TVL sequence, we calculate how long each fund stays in the staking pool. Hence, the TVL sequence for address \emph{0x014…c1} will be $[V_1, V_2, V_3, \cdots, V_n]$ with corresponding duration list $[t_2-t_1, t_3-t_2, \cdots, t_{n+1}-t_n]$, where $V_n=v_1+v_2+\cdots+v_n$ and $t_{n+1}$ represents the cut-off date for data collection (February 25, 2022).
Finally, we compute the weighted average of TVL for each address:
\begin{equation}
    \mbox{Weighted Average of TVL}=\frac{\sum_{i=1}^n V_i\cdot (t_{i+1}-t_i)}{t_{n+1}-t_1}
\end{equation}

\subsection{Clustering}
SOM \cite{vesanto2000clustering} is an unsupervised artificial neural network that applies a competitive learning strategy that neurons are activated under a mutually exclusive principle. This competition result can be achieved by having transverse inhibitory connections (negative feedback paths) between neurons so that neurons are forced to reorganize themselves. A nearest neighbor function is used to maintain the topology of the input space, which means that a two-dimensional map containing the relative distances between data and adjacent samples in the input space are mapped to adjacent output neurons. Compared with the traditional clustering method, it can keep the topological structure unchanged, and the clustering center formed can be mapped to a surface.
\begin{figure}[htbp]
	\centering
    \includegraphics[width=0.8\columnwidth]{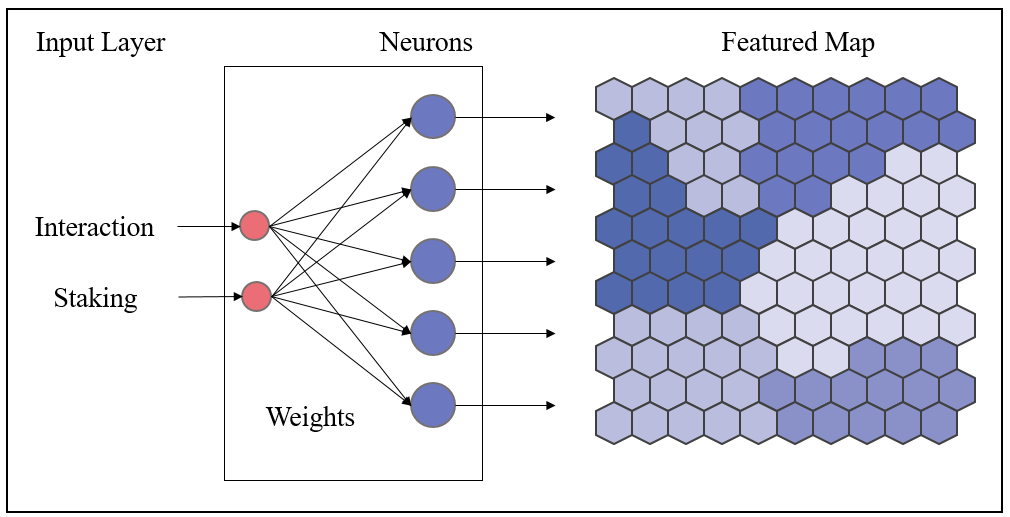}
    \caption{SOM’s Architecture}
    \label{fig:SOM}
\end{figure} 

As is shown in Figure \ref{fig:SOM}, we use the number of interaction times and the weighted average of TVL for each address as the input.
We traverse learning ratio and \emph{sigma} that controls the competitiveness of the neurons, finding a decent result with low quantization error where \emph{sigma} = 0.1 and learning ratio = 0.5.

\section{Analysis and Results}
This section finds out the user activity in blockchain games, analyzes player behavior, and discusses the impact of staking mainly after the official release of Aavegotchi from March 2, 2021, to February 25, 2022. 
We firstly present the activity of unique players and divide the game into five stages. Next, we analyze the functions called by users and the density distribution of duration of interaction and staking behavior. Besides, we investigate two specific users: those who only stake but do not interact and those who only interact but do not stake.
Then, we show the clustering result, discuss the player behavior in different clusters, and obtain the following insights: 
1) Player activity in Aavegotchi is supported by a few players with high-frequent gaming or high-volume staking.
2) In the short-term, staking can attract a large number of players, but cannot retain them. Well-designed gameplay should be the basis for the long-lasting retention of players.
3) The activity of staking is closely related to the market. 
As the price of GHST drops, many players will withdraw their staking funds, resulting in a decline in the number of players in the game.

 .

\subsection{The Five Stages of Aavegotchi Progress}
\begin{figure}[htbp]
	\centering
    \includegraphics[width=1\columnwidth]{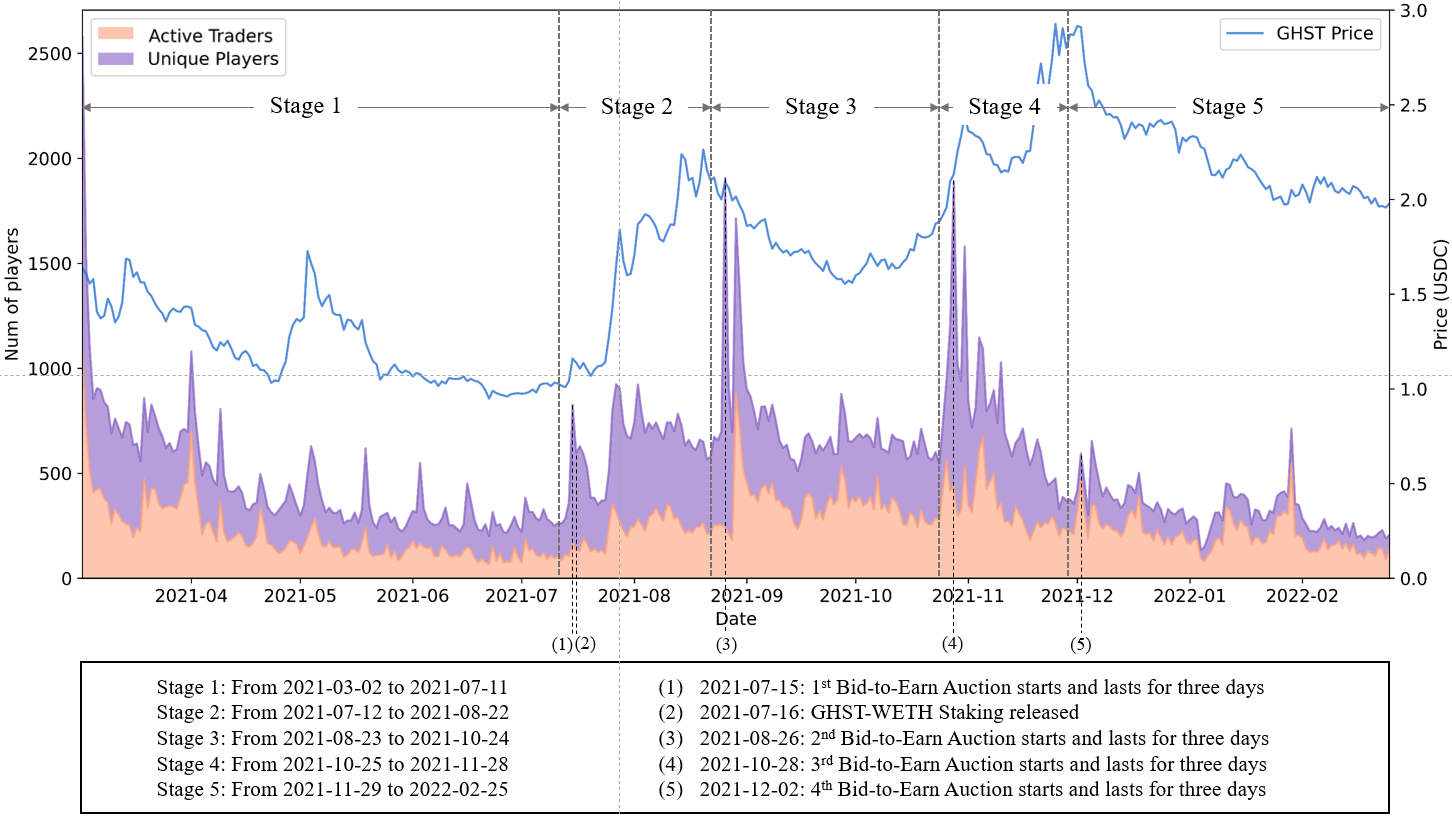}
    \caption{Player Status by Time}
    \label{fig:player_status}
\end{figure}

We extracted the participation of all addresses in the game and the marketplace and depicted the results in Figure \ref{fig:player_status}. We use an area chart (unstacked) to represent the activity of unique players (purple) and active traders (orange). GHST price is represented by the blue curve above. Selected special events of Aavegotchi are marked by vertical lines with annotation in the block under the chart. 
As we can see from the graph, special events significantly impact the activity of addresses.
Specifically, the number of daily unique players fluctuating declined until \emph{(1) 2021-07-15}, the 1\textsuperscript{st} bid-to-earn auction. One day after, GHST-ETH staking was released on \emph{(2) 2021-07-16}. After these two events, there has been a remarkable increment in the number of players and the price of GHST.
On \emph{(3) 2021-08-26}, the 2\textsuperscript{nd} bid-to-earn auction was launched for Gotchis and wearables,  leading to a peak in unique players and active traders. After this transient peak, the numbers recovered to a steady level. Even though the GHST price has slightly fallen, it stabilized at a relatively high level.
A similar phenomenon repeated itself at the 3\textsuperscript{rd} bid-to-earn auction for Gotchiverse land on \emph{(4) 2021-10-28}: a surge in user activity as well as the price of GHST, which gradually rose to a record high.
However, the identical incident did not happen a third time. The 4\textsuperscript{th} bid-to-earn auction for Gotchiverse land at \emph{(5) 2021-12-02} was not as effective as the previous two. Players started to drop out of the game and the market. At the same time, the price of GHST has gradually gone down.

Based on Aavegotchi's featured campaign and the activity records of players, the game can be divided into five stages: the introduction (Stage 1), the growth (Stage 2), the rise (Stage 3), the rush (Stage 4) and the decline (Stage 5), and the activity records of players and traders. Considering the publicity of the bid-to-earn events, the number of players will increase before the events start. Therefore, we set the splitting point of the stages three days before the event.

\begin{itemize}
    \item \textbf{The introduction stage (Stage 1)} ranges from March 2, 2021, to July 11, 2021, during which the game does not hold many events and gradually loses players.
    
    \item \textbf{The growth stage (Stage 2)} ranges from July 12, 2021, to August 22, 2021, when the 1\textsuperscript{st} bid-to-earn event and GHST-WETH staking are launched. Both the gameplay and the financial services of Aavegotchi are improved gradually.
    
    \item \textbf{The rise stage (Stage 3)} ranges from August 23, 2021 to October 24, 2021. Because of the success of the previous auction event, the auction is held once again for the Gotchis, strengthening the core of the gameplay.
    
    \item \textbf{The rush stage (Stage 4)} ranges from October 25, 2021 to November 28, 2021. Aavegotchi publishes the new project of Gotchiverse and starts the lands auction, enriching the game content and gameplay.
    
    \item \textbf{The decline stage (Stage 5)} ranges from November 29, 2021 to February 25, 2022. Aside from the second land auction, there are no other exceptional innovations or changes.
\end{itemize}


\subsection{Analysis of Functions Called}

We calculate the number of times five types of functions called: \emph{claim}, \emph{interact}, \emph{trading NFT}, \emph{trading item} and \emph{staking}. 

\begin{figure}[htbp]
	\centering
    \includegraphics[width=\columnwidth]{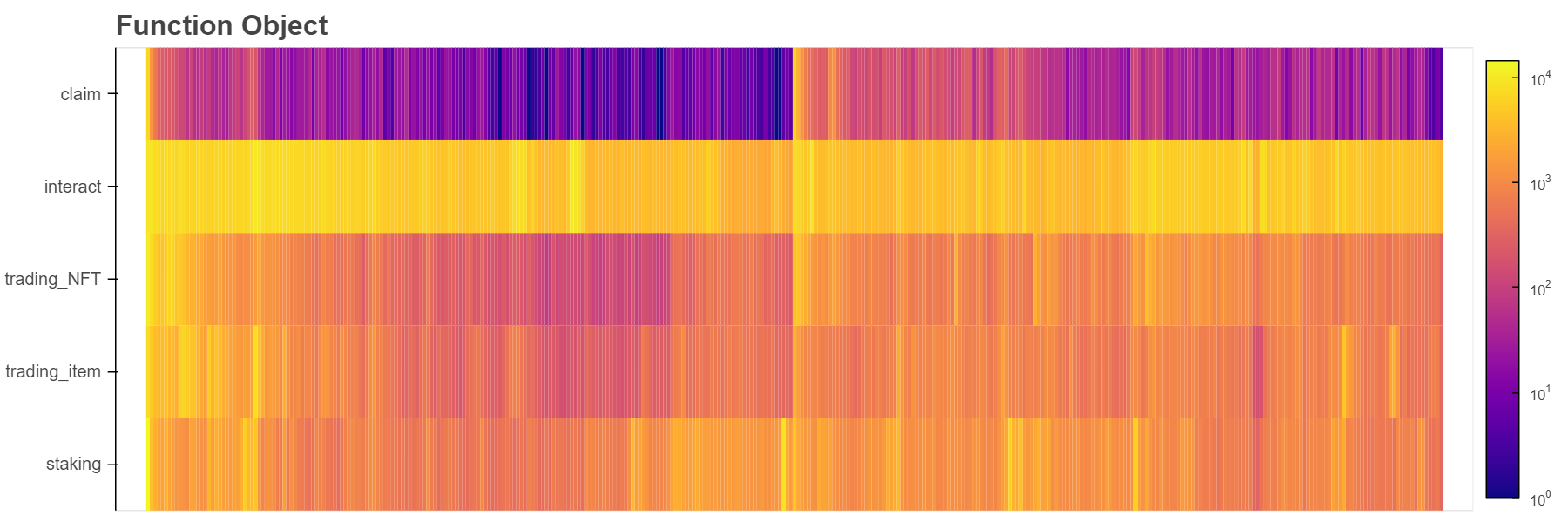}
    \caption{Number of Each Function Called Per Day}
    \label{fig:Num_Function_Called}
\end{figure}

	

As shown in Figure \ref{fig:Num_Function_Called}, each block represents a date, and the color of the block represents the number of calls. First, we focus on the whole picture that five types of functions have some degree of co-occurrence. A clear example of this is around September 2021, when there is a clear vertical highlighting band on the heat map, representing a significant increase in the number of function calls for all categories. This indicates the clear impact of the 2\textsuperscript{nd} bid-to-earn auction on the game activity. However, the 3\textsuperscript{rd} and 4\textsuperscript{th} bid-to-earn auctions around November and December 2021, the vertical light bands they formed on the heat map are then not that obvious.

On March 2, 2021, the call of \emph{staking} exploded to over ten thousand. In the introduction stage, the number of times \emph{staking} called gradually declines. We can observe that the number of calls increases during the growth stage. On the first day of the 2\textsuperscript{nd} bid-to-earn auction, there is a noticeable change in the color of the block for \emph{staking}. During the rise and rush stage, the number of times \emph{staking} called has generally remained in the thousands per day. Nevertheless, the calls have significantly increased at the 3\textsuperscript{rd} bid-to-earn auction and the 4\textsuperscript{th} bid-to-earn auction.
In the decline stage, there is a downward trend in the call of \emph{staking}. For the other four types of function, their trends in the number of calls are similar to \emph{staking}. 
However, it is worth mentioning that the call of the four types of function, \emph{claim}, \emph{interact}, \emph{trading NFT} and \emph{trading item}, grows appreciably after the 2\textsuperscript{nd} bid-to-earn auction, while the call of \emph{staking} booms at the first day of the 2\textsuperscript{nd} bid-to-earn auction. That's because the players can stake when participating in the auction. However, players get what they bid for after the auction, such as portals, wearables, and other items. Hence, the calls of \emph{claim}, \emph{interact}, \emph{trading NFT} and \emph{trading item} increase after the 2\textsuperscript{nd} bid-to-earn auction. Besides, since players can pet their Gotchis every 12 hours, \emph{interact} is called significantly more often than the other four.

Thus, from the trend of the five types of functions called, we can see that bid-to-earn activities that enable players to make money can only achieve short-term incentives. In other words, this money-making activity attracts the majority of speculators, who exit the market after a short period of profit. A few players are attracted by the gameplay and stay in the game, continuing to interact with Gotchis.


\subsection{Analysis of Density Distribution}
In this subsection, we apply kernel density estimation to analyze the duration of interaction and staking behavior for players. 
\begin{figure}[htbp]
	\centering
    \includegraphics[width=0.9\columnwidth]{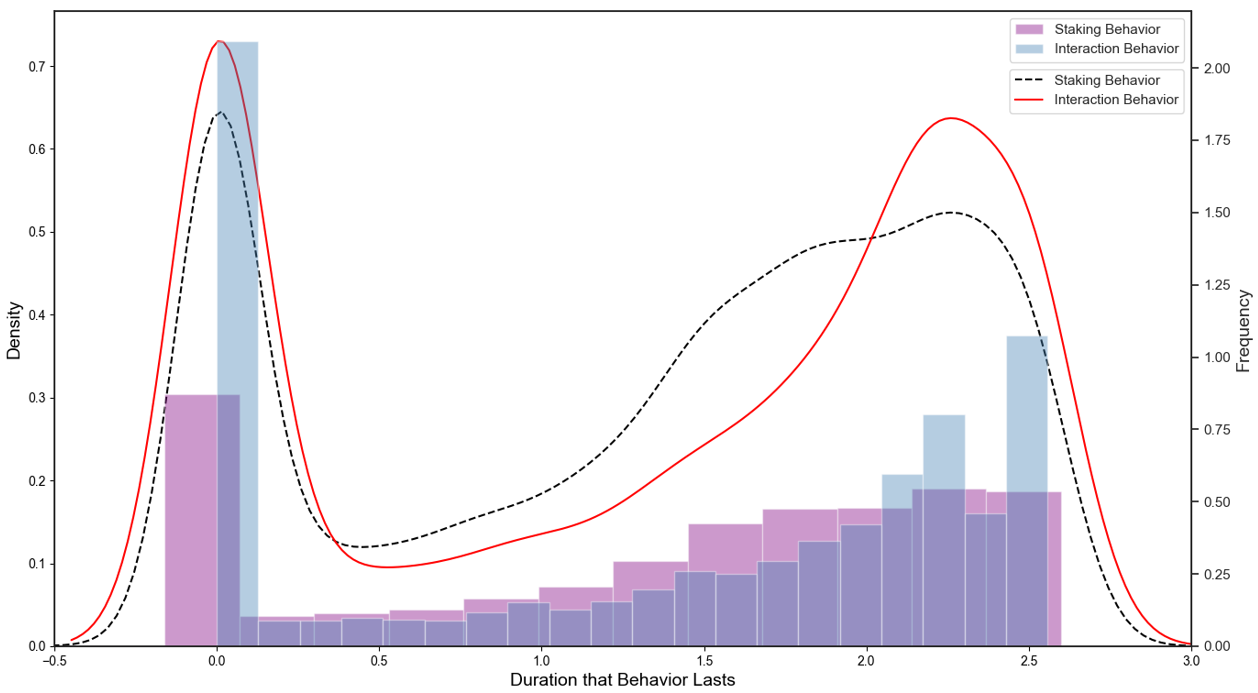}
    \caption{Logged Kernel Density Distribution of Duration of Interaction and Staking behavior}
    \label{fig:last_density}
\end{figure}

\begin{table*}[htbp]
    \centering
    \caption{The Number and Ratio of Specific Addresses in Each Stage}
    \label{tab:user_stage}
    \begin{tabular}{cccccccccc}
    \toprule
	\multirow{3}*{Stage} & \multirow{3}*{\makecell[c]{\# Address \\(Total)}} & \multicolumn{4}{c}{Stake-only Addresses} & \multicolumn{4}{c}{Interact-only Addresses} \\
	\cmidrule(lr){3-6}\cmidrule(lr){7-10}
	& & \multicolumn{2}{c}{In the Current Stage}& 
	\multicolumn{2}{c}{Remain in the Next Stage}&
	\multicolumn{2}{c}{In the Current Stage}& 
	\multicolumn{2}{c}{Remain in the Next Stage}\\
	\cmidrule(lr){3-4}\cmidrule(lr){5-6}\cmidrule(lr){7-8}\cmidrule(lr){9-10}
	& & \# Address &\% Address & \# Address &\% Address& \# Address &\% Address &\# Address &\% Address\\
	\midrule
	 1 & 10,760 & 5,702 & 52.99\% & 857 & 15.03\% & 188 & 1.75\% & 39 & 20.74\%\\
	 2 & 8,467 &  4,881 & 57.65\% & 2,146 & 43.97\% & 927 & 10.95\% & 741 & 79.94\%\\
	 3 & 10,570 & 4,936 & 46.71\% & 1,480 & 29.98\% & 860 & 8.14\% & 555 & 64.53\%\\
	 4 & 10,077 &  5,148 & 51.09\% & 1,897 & 36.85\% & 984 & 9.76\% & 788 & 80.08\%\\
	 5 & 14,069 &  8,504 & 60.44\% & - & - & 1,190 & 8.46\% & - & -\\
	\bottomrule
    \end{tabular}
\end{table*}

 As shown in Figure \ref{fig:last_density}, we can observe that most players only remain in the game for only one day, regardless of the interaction or staking operation. This part of the addresses is called sideliners. They just came in to quickly experience the game or staking services, then left with no return.
There are several reasons to explain this phenomenon: 
1) Systematic problems in gameplay design and staking mechanism lead to irretrievable user churn.
2) Excessive and unaffordable in-game assets discourage the entry of new players, hinder older players from keeping up with the pace of game content development, and further lead to fragmentation among the player community.
3) The 'inflation' in the in-game economic ecosystem implicitly shrinks the rewards gained from staking. After losing the novelty, relatively unprofitable staking services struggle to absorb more users.


Fortunately for Aavegotchi, some users still have been interacting or staking for more than 100 days.
The user base of Aavegotchi varies widely, and the gameplay evaluation system applicable to traditional hardcore gamers may not be able to fully describe the entertainment and financial services needs of blockchain players. But as far as the data-level findings have proven, even without user surveys, the vast majority of users' behavior rates the Aavegotchi experience negatively.

\subsection{Analysis of Specific Users in Stages}
In this section, we analyze the behavior of two specific categories of addresses: (1) stake-only addresses that only stake but do not interact, and (2) interact-only addresses that only interact but do not stake. 
As shown in Table \ref{tab:user_stage}, we count the number of addresses in each stage, the number and proportion of the two categories of users in each stage, and their activity data in the next stage, respectively.

Firstly, we discuss the stake-only addresses.
As shown in Table \ref{tab:user_stage}, stake-only addresses account for a large proportion that exceeds over 50\% in 4 out of 5 stages, which could be concluded as staking may be a considerable effective way to encourage user engagement.
It is worth noting that the GHST-WETH staking pool is opened to the public in the growth stage and offers more generous participation rewards, directly attracting more newcomers, making stake-only addresses occupy a much more significant percentage in the growth stage than the introduction stage. In addition, stake-only addresses account for a smaller portion in the rise stage and rush stage.
However, in the decline stage, stake-only addresses exceed 60\%.
This phenomenon is because, in the early stages, many arbitrageurs in a free market have led to high prices for Gotchis that are unaffordable so that the latecomers can only stake.
We also notice that the churn rate of this category is very high. 
For example, in the introduction stage, only 15.0\% of addresses remained in the next stage. This can be explained as the staking mechanism was not perfect until the GHST-WETH staking pool helped improve the retention rate. On the whole, staking rewards can motivate users to participate but cannot retain them long-term.

As for interact-only addresses, this category of addresses accounts for a tiny proportion in each stage. In the introduction stage, due to the imperfect gameplay, which is a common problem for most blockchain games, the number and the retention rate of interact-only addresses stay considerably low. As gameplay was improved and innovative activities such as bid-to-earn auctions are introduced in the upcoming stages, more players that can be categorized in interact-only addresses are attracted. Even more, the retention rate of this group is over 60\%. Based on the behavior of these addresses, improvement of the gameplay is clearly an effective way to hold users over the long run, proving that for blockchain games, having compelling gameplay is much more important than adding financial factors.


\subsection{Cluster Result}
In this subsection, we apply SOM to analyze the player behavior and summarise five clusters, named \emph{Lazy Player and Heavy Staker (Cluster 1)}, \emph{Crazy Player and Middle Staker (Cluster 2)}, \emph{Moderate Player and Light Staker (Cluster 3)}, \emph{Lazy Player and Light Staker (Cluster 4)} and \emph{Dispensable Player and Dispensable Staker (Cluster 5)}.

\begin{figure}[htbp]
	\centering
	\subfigure[Logged Number of Interaction for Cluster Result]{
		\begin{minipage}{0.47\columnwidth} 
            \includegraphics[width=\textwidth]{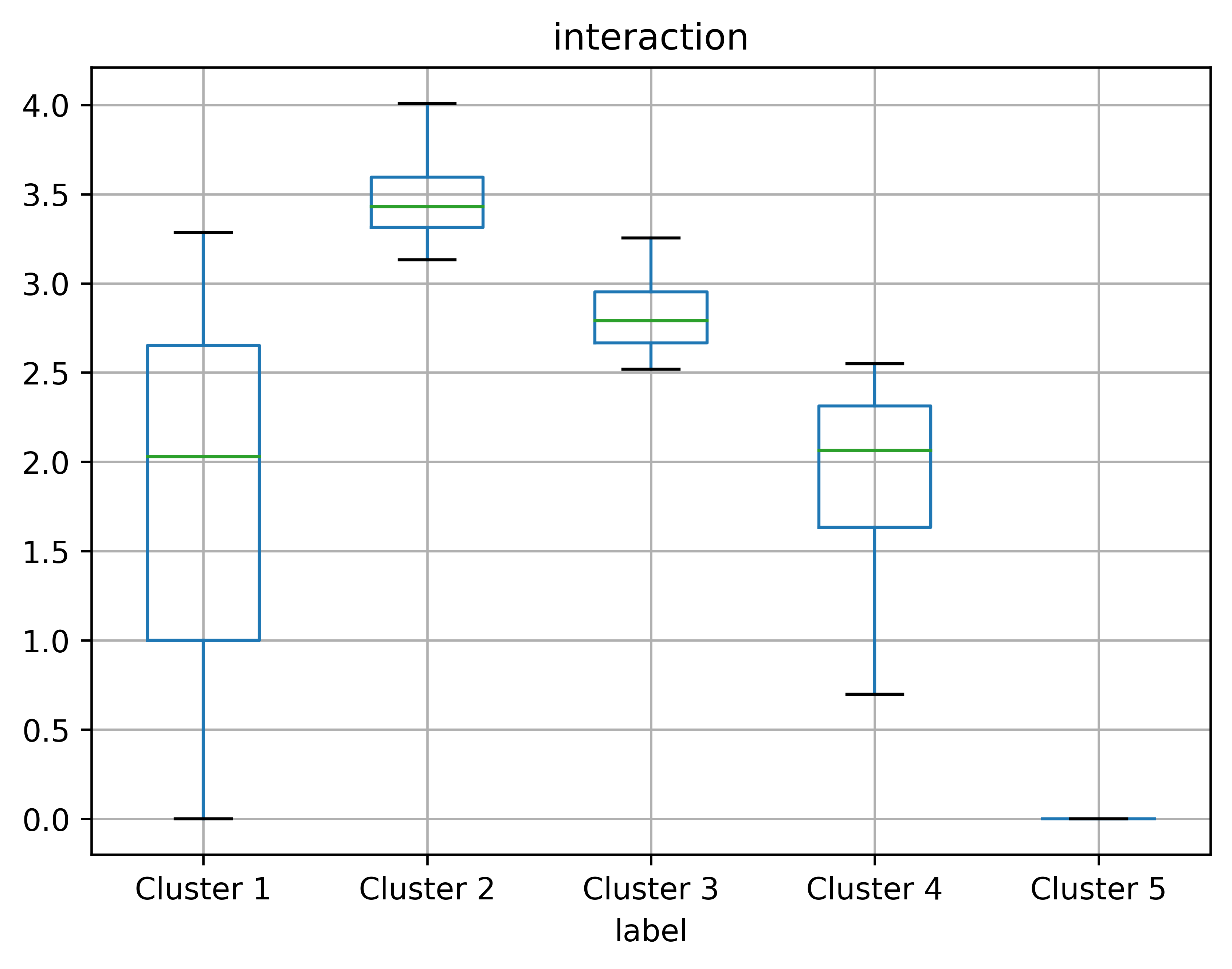} \\
            \label{fig:boxplot_interaction}
		\end{minipage}
	}
	\subfigure[Logged Weighted Average of TVL of Staking for Cluster Result]{
		\begin{minipage}{0.47\columnwidth}
			\includegraphics[width=\textwidth]{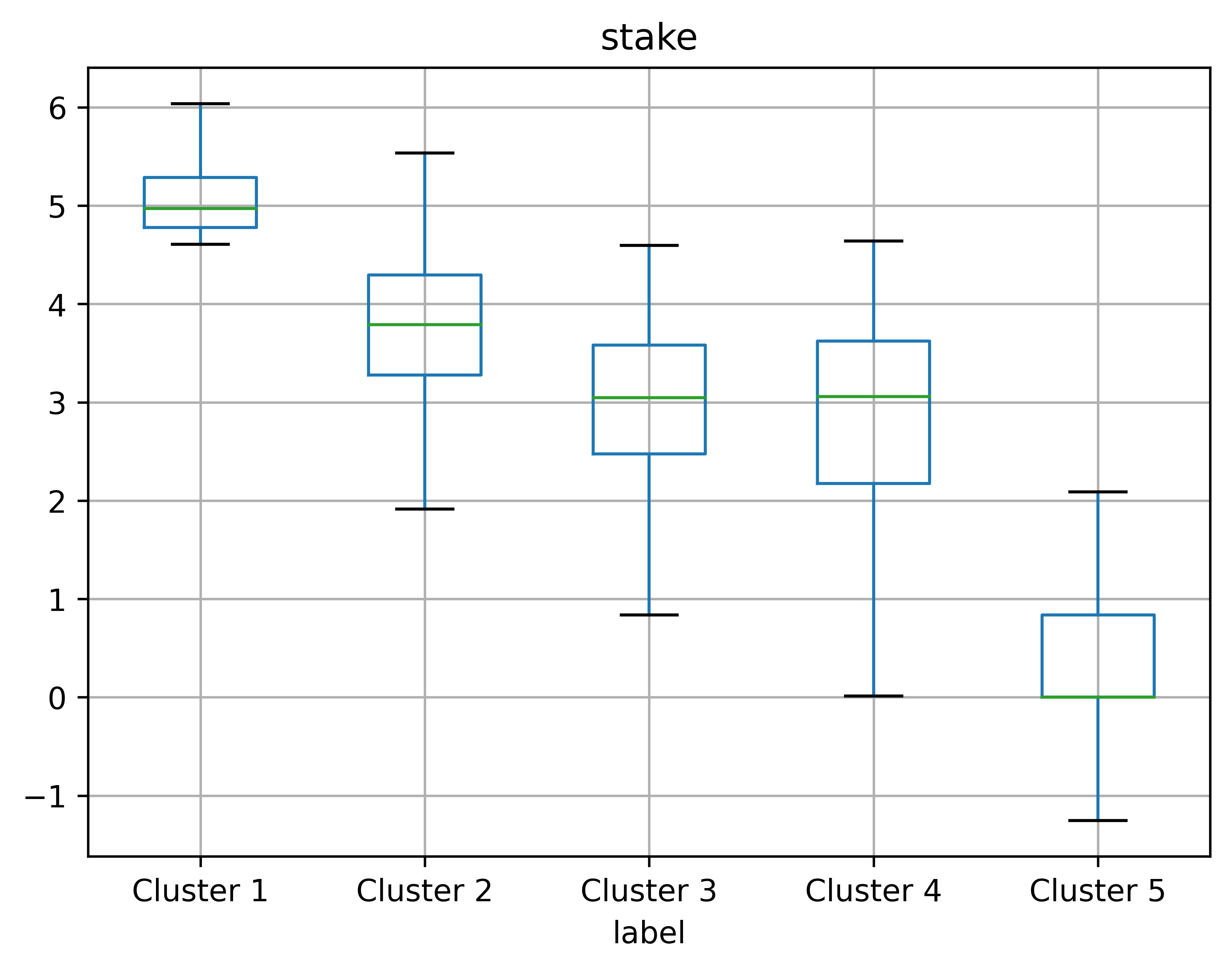}\\
			\label{fig:boxplot_stake}
		\end{minipage}
	}
	\caption{Logged Number of Interaction and Weighted Average of TVL of Staking for Cluster Result}
    \label{fig:cluster_boxplot}
\end{figure}

As shown in Figure \ref{fig:cluster_boxplot}, the descriptive data of the number of interactions and the weighted average of TVL of the groups are presented in the form of the boxplot.

\textbf{\emph{Lazy Players and Heavy Stakers (Cluster 1)}} refer to the addresses that interact little with the game but heavily invest in the staking pools, accounting for 0.50\% of total addresses.
From Figure \ref{fig:boxplot_interaction}, these addresses do not interact with the game frequently, and there is even a part of them that does not interact with the game at all.
However, from Figure \ref{fig:boxplot_stake}, these addresses have the largest weighted average of TVL in staking pools among the five clusters, providing a tremendous amount of liquidity that takes up more than half of the TVL of all.
For example, the address \emph{0x58f...98} locked total value worth over \$1,593,031 in the staking pool. 
The rich are not appear to be attracted by the gameplay but rather interested in staking in order to earn rewards quickly.

\textbf{\emph{Crazy Players and Middle Stakers (Cluster 2)}} represent the addresses that 
frequently interact with Gotchis and stake a relatively large amount of capital, taking up 0.94\% of all.
From Figure \ref{fig:boxplot_interaction}, this group of users is the most enthusiastic of all groups to participate in the game itself, to the extent that up to February 25, 2022, each user in this group interacted more than 1,000 times. Moreover, about 10\% of these users own more than 20 Gotchis.
Besides, from Figure \ref{fig:boxplot_stake}, 
this group staked quite a large amount of money with horizontal comparison. More than 80\% of them staked more than \$1,000.
This segment of users is more likely to be called real players, those who have passion in Gotchis, than \emph{Cluster 1} who are clearly motivated by profit. They bring more actual activity to the game and have a reason to be actively involved in discussions on the community and development of the game.

\textbf{\emph{Moderate Players and Light Stakers (Cluster 3)}} are addresses that 
are moderately involved in the game and staking compared to other groups, as shown in Figure \ref{fig:cluster_boxplot},
which accounts for 4.48\%.
These players' activities are relatively balanced from the aspect that they have positive interaction with Gotchi and can also provide a certain amount of liquidity in staking. This group of users can be considered fans of gameplay who acknowledge the exclusive design of Aavegotchi. However, they are not able to invest a tremendous amount of funds as \emph{Cluster 1} or leisure time as \emph{Cluster 2}. They don't stand out statistically.

\textbf{\emph{Lazy Players and Light Stakers (Cluster 4)}} are addresses that do not interact frequently but provide a certain large amount of liquidity, which take up 9.95\% of total addresses.
The main difference with \emph{Cluster 3} is that this group of users is not as keen to participate in the game in comparison.
This could be explained by the fact that Aavegotchi's gameplay did not appeal to this group or that their enthusiasm for gaming itself was limited to a moderately low level of engagement similar to \emph{Cluster 1}.

\textbf{\emph{Dispensable Players and Dispensable Stakers (Cluster 5)}}
refer to the addresses that lack enthusiasm for participation in neither gameplay nor staking, which accounts for 84.12\% of all.
Shown in Figure \ref{fig:cluster_boxplot}, the users in \emph{Cluster 5} rarely interact with the game. According to unlogged raw data, around 72.5\% of this group did not interact at all. Their staking amount is similarly unimpressive. If we look at the aspect of interaction, they may fit the characteristics of some addresses in \emph{Cluster 1}, but the difference is that they are cautiously reluctant to invest large amounts of money, motivated by novelty, or are simply conservative in making an attempt to invest in a project of combining game and finance.
We summarize the characteristics and motivations of this group as follows: 
1) They are users who are huge in number but cannot provide daily activity records for Aavegotchi or provide buzz for discussion on social media. The motivation of this part of participants should not be interested in the game itself but more likely to be curiosity-driven. 2)
They are similar to individual investors in the stock market, holding a small amount of capital and preferring to invest limited resources in projects with higher returns, given the volatile market of blockchain.
\begin{table}[htbp]
    \centering
    \caption{Number of Addresses for Engaged Times}
    \label{tab:Engaged_Times}
    \begin{tabular}{ccc}
    \toprule
    Engaged Times & \# Address & \% Addresses\\
    \midrule
    1 &  20,604&65.46\%\\
    2 &  4,771&15.16\%\\
    3 &  2,455&7.8\%\\
    4 &  1,895&6.02\%\\
    5 &  1,738&5.52\%\\
	\bottomrule
    \end{tabular}
\end{table}

\subsection{Analysis of Clusters in Different Stages}
In this section, we will discuss the difference among five clusters in different stages in terms of the interaction and staking behavior.
\subsubsection{The Distribution of Clusters by Engaged Times}
From Table \ref{tab:Engaged_Times}, there are 20,604 addresses, which account for over 50\% of all, that only participate in a single stage. Very few addresses went through the whole process. The game design of Aavegotchi is at issue in general.

\begin{table*}[htbp]
    \centering
    \caption{The Participation and Quit of Clusters in Interaction by Stages}
    \label{tab:Participation Interaction}
    \begin{tabular}{cccccccccccc}
    \toprule
	\multirow{2}*{\makecell[c]{Cluster\\ Label}} & 
	\multirow{2}*{\makecell[c]{\# Total \\ Addresses}}&
	\multicolumn{5}{c}{\# Addresses First Enter to the Aavegotchi}&
	\multirow{2}*{\makecell[c]{\# Full Participation\\ Addresses}}&
	\multirow{2}*{\makecell[c]{\# Return \\Addresses}}\\
	\cmidrule(lr){3-7}
	& &Stage 1 & Stage 2 & Stage 3 & Stage 4 & Stage 5 \\
	\midrule
	 1 &131 &60 &  28 & 16 &  17 &  10 & 28 & 19\\
	 2 &261 &241 &  12 &  5 &  2 &  1& 211 & 7 \\
	 3 &1,371 &988 & 118 &  200 &  38 &  27& 750 & 89 \\
	 4 &2,665 &1,658 &  416 &  854 & 373 &  364 & 396 & 491\\
	 5 &6,706 &2,105 &  1,032 &  1,500 &  888 &  1,181 & 11 & 445\\
	\bottomrule
    \end{tabular}
\end{table*}
\begin{figure}[htbp]
	\centering
    \includegraphics[width=0.91\columnwidth]{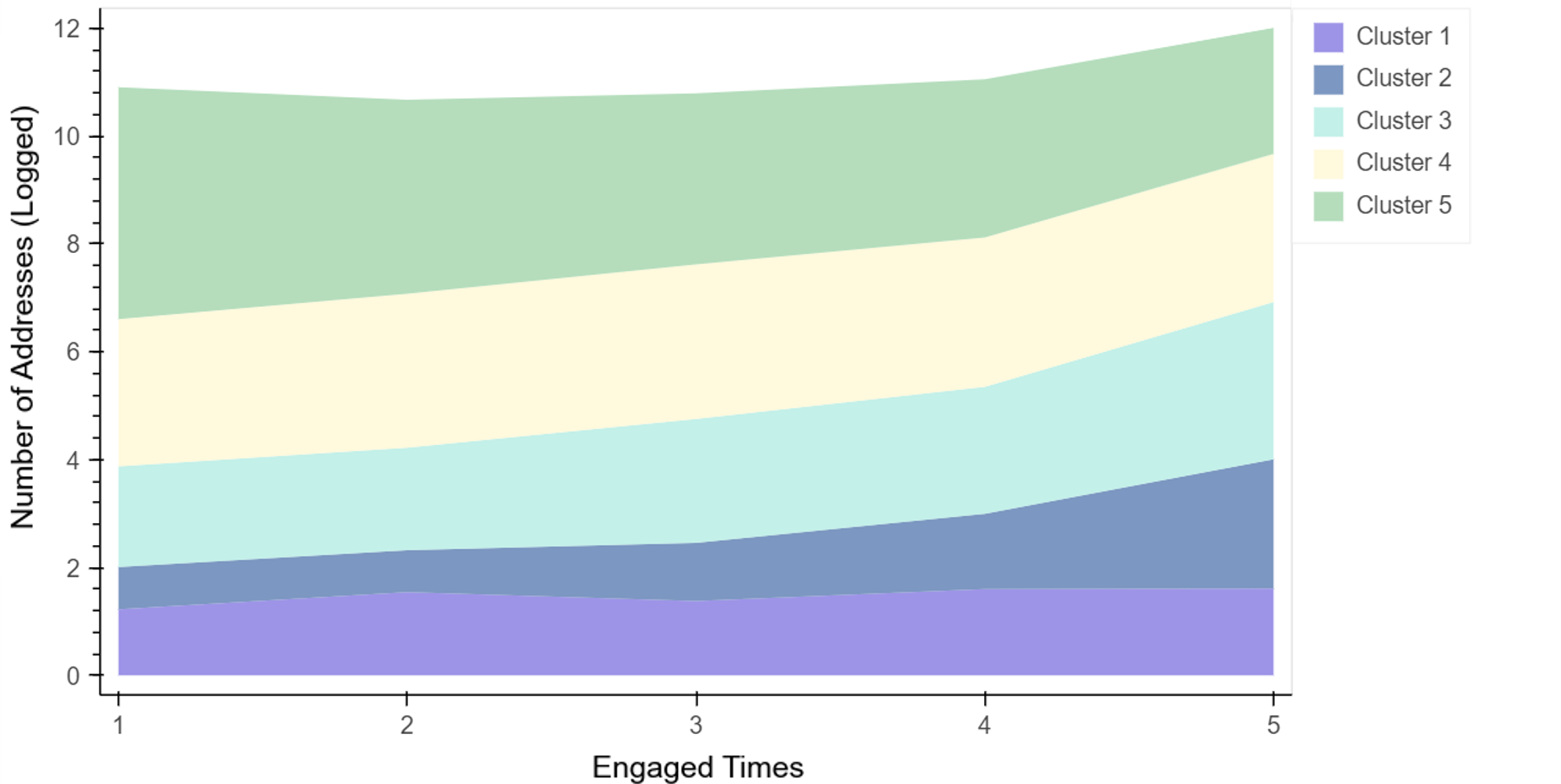}
    \caption{Logged Distribution of Clusters by Engaged Times}
    \label{fig:clusters_times}
\end{figure} 
We plot the distribution of address numbers in clusters by the count of stages they engaged in Figure \ref{fig:clusters_times}. The result shows that addresses in most clusters are quite evenly distributed across the number of stages experienced, but those clusters with more interactions, such as \emph{Cluster 2} and \emph{ Cluster 3}, have a higher proportion of more than 50\% addresses that experienced all five stages. The durability of user engagement is positively related to the level of interaction. Compared to \emph{Cluster 3}, \emph{Cluster 2} has more staking, which could be the reason behind it having more distribution in terms of fully engaged addresses. This fact may indicate that participating in both gaming and staking has a positive effect on user retention.

\begin{figure}[htbp]
	\centering
    \includegraphics[width=0.88\columnwidth]{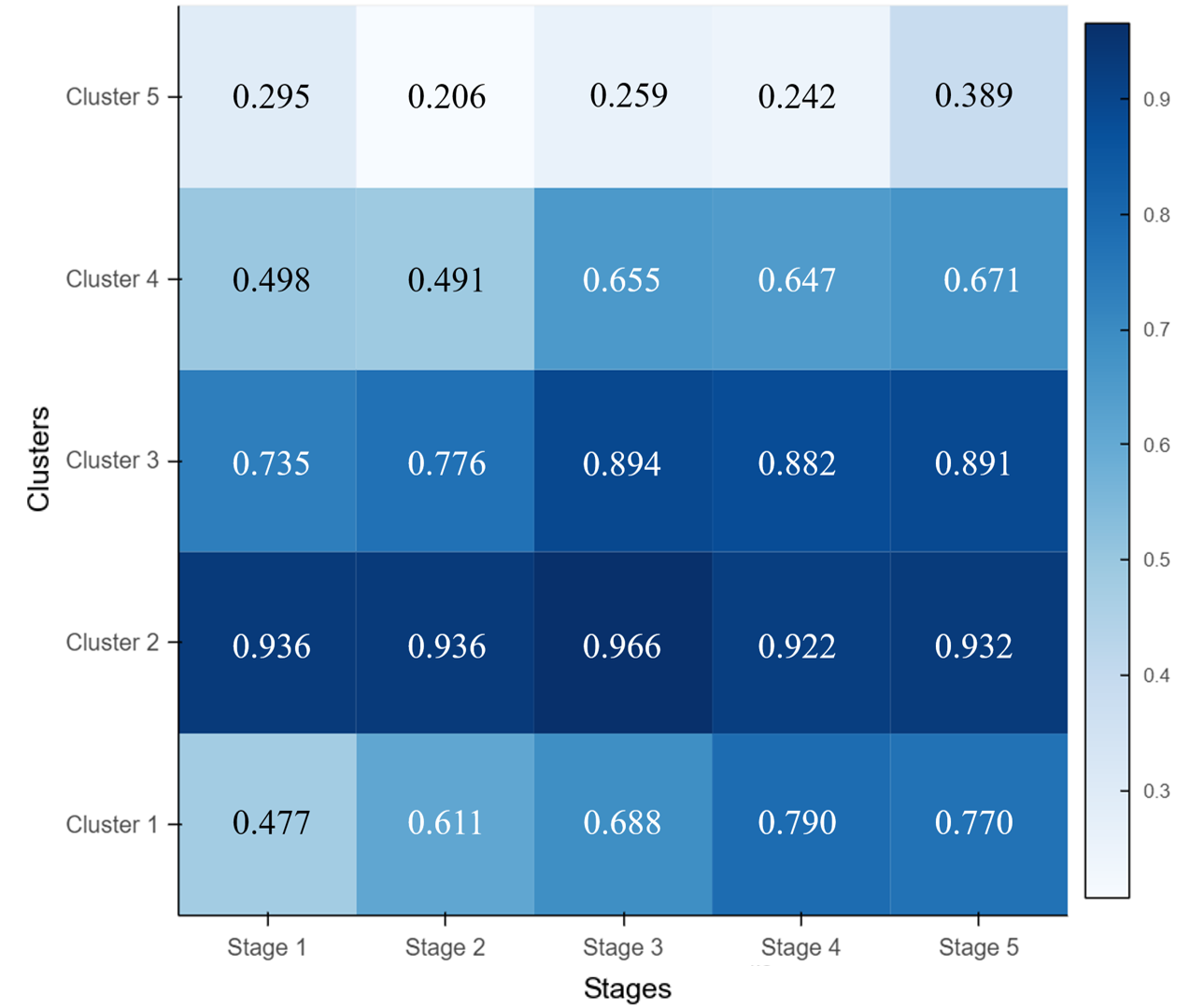}
    \caption{Ratio of Address Engaged in Stages by Clusters}
    \label{fig:Radio_Address}
\end{figure} 
\subsubsection{Ratio of Address Engaged in Stages by Clusters}
Figure \ref{fig:Radio_Address} shows the percentage of addresses of different clusters participating in different stages to the number of addresses in the corresponding cluster.
New-comers from \emph{Lazy Players and Heavy Stakers (Cluster 1)} kept flowing in until Stage 5, in which the largest increment happened in Stage 2, recalling that the GHST-WETH staking pool with high rewards was released at that time. However, in Stage 5, the rewards from the staking decreased as the TVL in the pool increased, causing withdrawal from staking.
\emph{Crazy Players and Middle Stakers (Cluster 2)} and \emph{Moderate Players and Light Stakers (Cluster 3)} are similar. 
They maintain a high percentage of addresses historically recorded engaged in all stages. The number of two clusters increases significantly in Stage 3 because the 1\textsuperscript{st} bid-to-earn auction was held.

\subsubsection{The Interaction of Clusters by Stages}
In this part, we will analyze the participation and quit of clusters by stages from interaction. Table \ref{tab:Participation Interaction} shows the number of addresses that join the Aavegotchi for game interaction by stages.

Judging from the number of participants in Stage 1, we found a remarkable \emph{preemptive advantage}. If we focus on the groups of users who contribute the most to the game and stay involved the longest, the majority of \emph{Cluster 2} and \emph{Cluster} 3 joined since the early days of the game. They can purchase assets at a lower price and have a more significant advantage over the latecomers as the game develops with a rising asset price: the pioneers never lose money. This advantage allows the pioneers to continuously invest resources, including time and money, at a lower risk and transforms into a strong competitive power in the NFT market, crowding out the ability of latecomers to purchase and participate. Above all, they could face asset inflation at a later stage comfortably.

However, for the marginal \emph{Cluster 4} and \emph{Cluster 5}, although a considerable portion of their users also belongs to the pioneers, it seems that they are not active enough to participate in neither gameplay nor staking from the statistical findings of Figure \ref{fig:cluster_boxplot}. Throughout the development of the game, later campaigns and game policies, such as Stage 2 and Stage 3, have much more attractive and motivating effects on the marginal groups than \emph{Cluster 2} and \emph{Cluster 3}, which indicates that the urgent needs for blockchain games should revolve around how to convert the light users attracted into long-term users without harming the pioneers' interests. This may require 1) a more reasonable token distribution model, 2) a planned regulation policy for the NFT market, 3) adequate benefits and protection zones for new users, 3) the improvement of distribution in the value of Gotchis, and lowering the threshold for acquisition and cultivation so that Aavegotchi can run into a long-term business with better gameplay.
\begin{figure*}[htbp]
	\centering
    \includegraphics[width=1\textwidth]{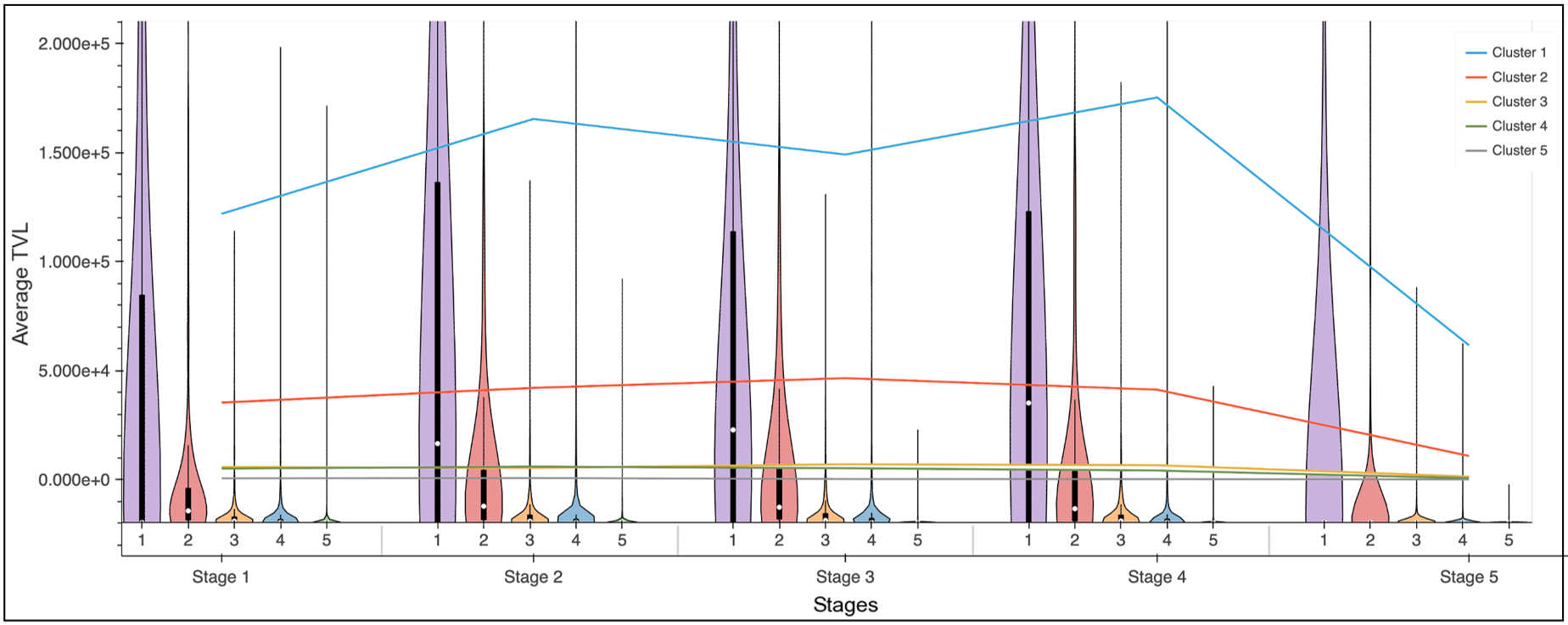}
   \caption{TVL Distribution of Clusters by Stages}
    \label{fig:violin_chart}
\end{figure*} 
\subsubsection{The Staking of Clusters by Stages}
Figure \ref{fig:violin_chart} recorded the average TVL of each cluster in curves, with their TVL distribution shown in the form of a violin graph. The stage increases along the x-axis, and the number from 1 to 5 above each stage represents the label of the corresponding cluster.

Intuitively, the TVL of the clusters differs significantly in value. Still, in terms of distribution, except for \emph{Cluster 1} which contains a relatively small number of addresses, the distribution of the other clusters is concentrated around a low TVL value with a long tail. Regardless of the stage, \emph{Cluster 1} distinctly dominates the total TVL supply of the game, while the difference between \emph{Cluster 3, 4, 5} is not as noticeable.

From a cross-sectional view, we can find that at Stage 5, a significant drop in all clusters occurred. We attribute this phenomenon to the impact of the overall market on user decisions. Recalling the stage division in Figure \ref{fig:player_status}, Stage 5 refers to the period from the end of November 2020 to the present. During this period, the price of Ether, which is a sign of the blockchain market, fell 45\%. Meanwhile, statistics about Aavegotchi show a gradual decline in the number of active addresses and GHST prices. Users naturally turn funds in financial products into stable assets to hedge risks when facing market shocks. This indicates that when blockchain games are trying to build up an in-game financial system connected to the whole blockchain market to enjoy greater potential benefits, they also face a considerable risk of embracing impacts. From Aavegotchi, the decrease in daily active users is perhaps the most unacceptable for a game. Whether in-game financial systems should be linked to the free market should be one of the most critical topics for game researchers when studying blockchain games.

After five stages, the opening of a new staking pool, the introduction of game events and auctions, will Aavegotchi's efforts result in more users participating in staking? If we look at Cluster 5, which has the most significant number of users and is the most representative in percentage, the answer is \emph{No}. In terms of distribution, the layout of Cluster 5 is gradually concentrated to 0 over time, which means that most users are less willing to participate in staking as the game activity progresses. In their case, a simple lack of available funds cannot be used as a reason for not participating in staking. They have the freedom of choice, and the reward of staking offered by Aavegotchi may not be competitive in terms of risk and rate of return compared to financial services provided by other organizations, which is even more so for latecomers who missed the preemptive advantage.

In general, the beautiful ideas of setting up an in-game financial system linked to the market, financing the players in a staking way to facilitate the development of the game to form a positive feedback loop do not achieve the goals they were designed for. Only a tiny number of preemptive users and large asset holders can be the beneficiaries of such activities. In contrast, most users are players who are gradually disappointed.

\section{Discussion}
In the following section, we discuss our findings and attempt to answer three research questions previously proposed in Section \ref{sec:intro}. 
\subsection*{Q1. What's the trend of player activity in blockchain games?}
We define the two types of operation for players: interaction behavior and staking behavior.

Regarding interaction behavior, players are usually more active in the early stages of a game.
The reason is that the price of the token and NFT is relatively low at the beginning of the game with low barriers to entry, which is acceptable for many players to interact with the game. The prices of both the token and NFT increase as the game develops. The number of players who are willing to interact with the game does not increase as the popularity of the game because most of them are blocked by affordable assets price. 

Staking behavior is similar to interaction behavior in the early stages. That's because the player who stakes early can receive more rewards. The rewards for staking decrease after more players join the staking, and therefore the activity level decline since it's less profitable. The activity level of staking behavior is also positively affected by new staking mechanisms, for example, new staking pools. The players will be attracted by the high rewards of the new pool at the beginning.

\subsection*{Q2. What attracts players to blockchain games, and why do they leave?}

Well-designed gameplay will attract users and retain them. Moreover, users who like the gameplay design can stay in the game for a long time without leaving.

Financial services can, to some extent, be attractive to players. From the data of Aavegotchi, we can see that about 50\% of users only stake but do not interact with the game, which reflects that most users are motivated by financial activities. In the blockchain world, there is a large gap between the number of assets users hold. Those with abundant assets can stake a large amount of capital to earn considerable rewards in a short time. For those with insufficient funds, staking provides a lower threshold than the NFT market. They can stake any amount of capital into the pools to earn rewards rather than buying expensive NFT to participate in the game. 
However, as we mentioned in the analysis section, the appeal of staking to the vast majority of general users is limited in the long run, and staking does not significantly impact game engagement. With the gradual decline of the staking rewards, the number of participants reduced considerably to burst the illusion of the early prosperity of the game, and the bursting of the bubble may lead to more serious adverse effects, such as NFT and other assets that linked to the game may face a substantial devaluation for losing demand. This devaluation is usually irreversible and will further lead to the collapse of the in-game economic system.

Besides, some special events or mechanisms that allow players to make money will also attract players to enter. A large proportion of these players are speculators who see lucrative activities. If there are no lasting benefits to attract this group of speculators, they will soon leave the game.

\subsection*{Q3. What should the design of blockchain games focus on? }
Blockchain games should be cautious about the in-game economic system and carefully consider to what extent it should connect to the overall blockchain market. At the same time, the need for better gameplay development should take precedence over the construction of financial services.

Gameplay is the key to user retention. Even though blockchain itself has an inseparable financial factor, as a blockchain game, we should not forget its essence as a game. However, many blockchain games currently on the market abuse the publicity gimmick of \emph{"To Earn"} and ignore the importance of gameplay, which creates a negative atmosphere of fast-paced and impetuousness for both developers and users.

Setting a reasonable and long-lasting in-game economic system is already a task that requires a lot of skill and experience to balance the numerical setting from player's income to asset's price. However, most small-volume blockchain games try to operate an in-game market that directly interfaces with the huge blockchain market, which by implication is the real-world currency, without the capability of controlling a decent in-game value design. The prevailing overprices and plunges in the blockchain market have led to extremely volatile asset prices, and these are the very reasons why users are pulling out of assets and thus losing interest.

Financial services offered by the blockchain games are similar to a coin with two sides. Positively, staking has a lower threshold than NFT trading, allowing users with a small amount of money to earn in-game rewards. In addition, staking can be considered attractive to new users in the short term. On the negative side, staking is not a sustainable solution. As time goes on, the yield decreases, and the churn rate continues to grow. 

In brief, blockchain needs a phenomenal game that implicitly incorporates the economic and financial features of blockchain and rivals traditional games in terms of gameplay in order to rebrand a decent reputation. As for the in-game economic and financial system, developers should optimize the token issuance model and carefully control the degree of its connection with the blockchain free market or, in the context of a free market, how to figure out a way to effectively stabilize the prices of game-related assets, give more space to light players, and avoid blockchain games from becoming entertainment for minority elites, which is essential to the further realization of Metaverse.



\section{Conclusion}
In this paper, we attempt to identify the user activity level of blockchain games and the impact of financial activities.
We choose Aavegotchi as a case and collect all addresses from February 2, 2021, to February 25, 2022. After extracting the feature from two aspects, interaction and staking, we use SOM to compare player behaviors in different stages.
We then visualized and analyzed the player profile and cluster result.
We observe that player activity is not high in either gameplay or financial activities.
Very few users provide most of the activity for financial activities. Similarly, a small percentage of players interact with the game frequently, and more than half are dispensable.
Besides, we conclude that financial activities are indispensable for blockchain games and play a unique role in attracting players and providing a way for players to engage with the game.
However, financial incentives tend to create the illusion of a boom in game players. Once the market fades, players will drop out of the game.
Finally, blockchain games cannot ignore the improvement of gameplay, which is crucial to the long-term retention of players.
\begin{acks}
This work was supported by Shenzhen Science and Technology Program (Grant No. JCYJ20210324124205016), in part by y2z Ventures.
\end{acks}

\bibliographystyle{ACM-Reference-Format}
\bibliography{sample-manuscript}

\appendix

\end{document}